\documentclass[a4paper,twocolumn,showpacs,superscriptaddress,floatfix]{revtex4-1}
\usepackage{graphicx}
\usepackage{epsf}
\usepackage{epsfig}
\usepackage{amssymb,amsmath}
\usepackage[usenames]{color}
\usepackage{amssymb}
\usepackage{times}

\begin{document}

\title{Quantum Supersymmetric Cosmology and its Hidden Kac-Moody Structure}

\author{Thibault \surname{Damour}}
\affiliation{Institut des Hautes \'Etudes Scientifiques, 91440 Bures-sur-Yvette, France}

\author{Philippe \surname{Spindel}}
\affiliation{M\'ecanique et gravitation, Universit\'e de Mons, 7000 Mons, Belgique}

\begin{abstract}
We study the quantum dynamics of a supersymmetric squashed three-sphere by dimensionally reducing (to one timelike dimension) the action of $D=4$ simple supergravity for an $SO(3)$-homogeneous (Bianchi IX) cosmological model. The quantization of the homogeneous gravitino field leads to a 64-dimensional fermionic Hilbert space. The algebra of the supersymmetry constraints and of the Hamiltonian one is found to close. One finds that the quantum Hamiltonian is built from operators that generate a 64-dimensional representation of the (infinite-dimensional) maximally compact sub-algebra of the rank-3 hyperbolic Kac-Moody algebra $AE_3$. Some exponentials of these operators generate a spinorial extension of the Weyl group of $AE_3$ which describe (in the small wavelength limit) the chaotic quantum evolution of the universe near the cosmological singularity.
\end{abstract}


\maketitle

One of the key challenges of gravitational physics is to understand the fate of space-time at spacelike (cosmological) singularities, such as the big bang singularity that gave birth to our universe. A novel way of attacking this problem has been suggested a few years ago via a conjectured {\it correspondence} between various supergravity theories and the dynamics of a spinning massless particle on an infinite-dimensional Kac-Moody coset space~\cite{Damour:2002cu,Damour:2005zs,de Buyl:2005mt,Damour:2006xu}. Evidence for such a supergravity/Kac-Moody link emerged through the study \`a la Belinskii-Khalatnikov-Lifshitz (BKL)~\cite{Belinsky:1970ew} of the structure of cosmological singularities in string theory and supergravity, in spacetime dimensions $4 \leq D \leq 11$~\cite{Damour:2000hv,Damour:2001sa,Damour:2002et}.  For instance,  the well-known BKL oscillatory behavior~\cite{Belinsky:1970ew}  of the diagonal components of a generic, inhomogeneous Einsteinian metric in $D=4$ was found to be equivalent to a billiard motion within the Weyl chamber of the rank-3 hyperbolic Kac-Moody algebra $AE_3$~\cite{Damour:2001sa}. Similarly,
the generic BKL-like dynamics of the bosonic sector of maximal supergravity (considered either in $D=11$, or, after dimensional reduction, in $4 \leq D \leq 10$) leads to a chaotic billiard motion within the Weyl chamber of the rank-10 hyperbolic Kac-Moody algebra $E_{10}$~\cite{Damour:2000hv}.  The hidden r\^ole of $E_{10}$ in the dynamics of maximal supergravity was confirmed to higher-approximations (up to the third level) in the gradient expansion $\partial_x \ll \partial_T$ of  its bosonic sector~\cite{Damour:2002cu}. In addition, the  study of the fermionic sector of supergravity theories has exhibited a related r\^ole of Kac-Moody algebras. At leading order in the gradient expansion of the gravitino field $\psi_{\mu}$,  the dynamics of $\psi_{\mu}$ at each spatial point was found to be given by  parallel transport with respect to a (bosonic-induced) connection $Q$ taking values within the ``compact'' sub-algebra of the corresponding bosonic Kac-Moody algebra: say $K(AE_3)$ for $D=4$ simple supergravity and $K(E_{10})$ for maximal supergravity~\cite{Damour:2005zs,de Buyl:2005mt,Damour:2006xu}. However,  the latter works considered only the terms {\it linear} in the gravitino, and, moreover, treated  $\psi_{\mu}$ as a ``classical'' (i.e. Grassman-valued) fermionic field.

The aim of this communication is to clarify the occurrence of hidden Kac-Moody structures in simple supergravity, within a setting which goes beyond previous work both by being {\it fully quantum}, and by taking completely into account the crucial {\it nonlinearities in the fermions} that allow supergravity to exist. On the other hand, our framework will simplify the cosmological dynamics by working within a supersymmetric minisuperspace model, namely a Bianchi IX one. Though the quantum theory of supersymmetric minisuperspace models  has attracted the interest of many authors~\cite{D'Eath:1993ki,D'Eath:1993up,Csordas:1995kd,Csordas:1995qy,Obregon:1998hb} , we shall give here, for the first time, a complete description of all the physical states of the supersymmetric Bianchi IX model.

Our formalism is a generalization of the formalism we used in Ref.~\cite{Damour:2011yk}  to study the quantum dynamics of Einstein-Dirac Bianchi universes. It differs from the formalisms used in previous works~\cite{D'Eath:1996at,VargasMoniz:2010rxa}  in describing the gravity degrees of freedom entirely in terms of the metric components $g_{\mu\nu}$. We use the symmetry properties of Bianchi models to uniquely determine a specific vielbein $h_{ \ \ \mu}^{\hat\alpha}$ (with $g_{\mu\nu} = \eta_{\hat\alpha\hat\beta} \, h_{ \ \ \mu}^{\hat\alpha} \, h_{ \ \ \nu}^{\hat\beta}$) as a local function of $g_{\mu\nu}$. In other words, we gauge-fix from the start the six extra degrees of freedom contained in $h_{ \ \ \mu}^{\hat\alpha}$ that could describe arbitrary local Lorentz rotations. This gauge-fixing of the local $SO(3,1)$ gauge symmetry eliminates the need of the usual formalisms~\cite{D'Eath:1996at,VargasMoniz:2010rxa} to impose the six local Lorentz constraints $J_{\hat\alpha\hat\beta} \approx 0$.  

We start from the metric describing a time-dependent, $SO(3)$-homogeneous triaxially squashed $3$-sphere,
\begin{eqnarray}
\label{eq1}
g_{\mu\nu} \, dx^{\mu} \, dx^{\nu} &= & 
- \, N^2 (t) dt^2  \\ &+ &g_{ab} (t) (\tau^a (x) + N^a(t) dt) (\tau^b (x) + N^b (t) dt) \, , \nonumber 
\end{eqnarray}
where the left-invariant one-forms $\tau^a (x) = \tau_i^a (x) \, dx^i$ (which only depend on spatial coordinates $x^i$) satisfy $d\tau^a = \frac12 \, C_{ \ \ bc}^a \, \tau^b \wedge \tau^c$ with the usual $SO(3)$ structure constants $C_{ \ \ bc}^a = \varepsilon_{abc}$. We then parametrize the metric $g_{ab}(t)$ in terms of three diagonal degrees of freedom $\beta^a (t)$, $a=1,2,3$ and of the three Euler angles $\varphi^a (t)$ describing the orthogonal matrix $S_{ \ \ b}^{\hat a} (\varphi^c)$ entering the Gauss decomposition of the symmetric matrix $g$: $g_{bc} = {\sum_a} \, e^{-2\beta^a} S_{ \ \ b}^{\hat a} \, S_{ \ \ c}^{\hat a}$. From these data, we then uniquely specify a vielbein coframe $\theta^{\hat\alpha} = h_{ \ \ \mu}^{\hat\alpha} \, dx^{\mu}$ as $\theta^{\hat 0} = N(t) dt$, $\theta^{\hat a} = {\sum_b} \, e^{-\beta^a (t)} S_{ \ \ b}^{\hat a} (\varphi^c(t))(\tau^b (x) + N^b(t) dt)$. The corresponding (time-dependent, $SO(3)$-homogeneous) gravitino field is described by its 16 vielbein components $\psi_{\hat\alpha}^A(t)$, where $\hat\alpha = 0,1,2,3$ is a four vector index linked to $\theta^{\hat\alpha}$, and where $A=1,2,3,4$ denotes a Majorana spinor index. Following previous work, it is convenient to replace $\psi_{\hat\alpha}^A (t)$ by the rescaled gravitino field $\Psi_{\hat\alpha}^A (t) := g^{1/4} \, \psi_{\hat\alpha}^A$ where $g^{1/4} = \exp \left(-\frac12 \, \beta^0 \right)$ (with $\beta^0 := \beta^1 + \beta^2 + \beta^3$). This  eliminates the couplings $\sim \dot \beta \, \psi$  in the action. Inserting these definitions in the supergravity action ${\mathcal L}$ \cite{Freedman:1976xh,Deser:1976eh}, and passing to its Hamiltonian version, in terms of the bosonic momenta $\pi_a \equiv \partial \, {\mathcal L} / \partial \, \dot\beta^a$ and $p_{w^a} =  \partial \, {\mathcal L} / \partial \, w^a$  (where $w^1 , w^2 , w^3$ denote the three independent angular velocity components $\sim \dot\varphi^a$ of the time-dependent rotation matrix $S$ : $(\dot S S^{-1})^{\hat a}_{ \ \ \hat b} = \varepsilon_{\hat a \hat b \hat c} \, w^{\hat c})$ leads to an Hamiltonian action of the form (we use units where $c=\hbar = 1$ and $8\pi \, G = V_3 = 16 \pi^2$, so as to absorb the volume $V_3$ of the  3-sphere, of curvature $1/4$, corresponding to $\beta^a=0$)
\begin{eqnarray}
\label{eq2}
{\mathcal L}_H &= &\pi_a \, \dot\beta^a + p_{w^a} \, w^a + \frac12 \, \bar\Psi_{\hat a} \, \gamma^{\hat a \hat 0 \hat b} \, \dot\Psi_{\hat b} \\
&+ &\bar\Psi'_{\hat 0} \, {\mathcal S} - \tilde N H - N^a H_a \, . \nonumber 
\end{eqnarray}
Here, we suppressed the spinor indices on $\Psi$ or $\gamma$, $\bar\Psi := i \Psi^{T} \gamma_{\hat 0}$ denotes a Majorana conjugate and $\tilde N \equiv N/\sqrt g = N \exp \beta^0$. We use a Majorana (i.e. real) representation of the four Dirac gamma matrices $\gamma^{\hat\alpha}$ (satisfying $\gamma^{\hat\alpha} \gamma^{\hat\beta} + \gamma^{\hat\beta} \gamma^{\hat\alpha} = 2 \, \eta^{\hat\alpha\hat\beta}$); see e.g.  Eq.~(4.6) in \cite{Damour:2011yk}.  
$\Psi'_{\hat 0}$ denotes the combination $\Psi'_{\hat 0} := \Psi_{\hat 0} - \gamma_{\hat 0} \, \gamma^{\hat a} \, \Psi_{\hat a}$. Eq.~(\ref{eq2}) exhibits the presence of three types of Lagrange multipliers appearing linearly in the action~: $\bar\Psi'^A_{\hat 0}$ (linked to local supersymmetry), $\tilde N$ (linked to time reparametrizations) and $N^a$ (linked to spatial diffeomorphisms). Their variations lead to three types of corresponding constraints: the four supersymmetry constraints ${\mathcal S}_A \approx 0$, the Hamiltonian constraint $H \approx 0$, and the diffeomorphism constraints $H_a \approx 0$.

We quantize the constrained dynamics, Eq.~(\ref{eq2}), by first reading off the (anti)commutation relations among the bosonic (fermionic) variables $\beta^a , \pi_a , \varphi^a, p_{\varphi^a} \sim p_{w^a} (\Psi_{\hat a}^A)$ from the kinetic terms in (\ref{eq2}). The quantization of the bosonic variables is conveniently done in a Schr\"odinger-like representation with the wave function of the universe taken as a function of the three logarithmic scale factors $\beta^a$ and the three Euler angles $\varphi^a$. Then $\hat\pi_a = -i \, \partial / \partial \, \beta^a$, $\hat p_{\varphi^a} = - i \, \partial / \partial \, \varphi^a$, together with the natural ordering of the $\hat p_{w^a}$'s as differential operators on the $SO(3)$ space (see \cite{Damour:2011yk}). The quantization of the gravitino operators $\hat\Psi_{\hat a}^A$ is simplified by introducing the new gravitino variables~\cite{Damour:2009zc} $\hat\Phi_A^a := \sum_B \gamma_{ \ \ AB}^{\hat a} \, \hat\Psi_{\hat a}^{B}$ (no summation on $\hat a$),  whose quantization conditions read  
\begin{equation}
\label{eq3}
\hat\Phi_A^a \, \hat\Phi_B^b + \hat\Phi_B^b \, \hat\Phi_A^a = G^{ab} \, \delta_{AB} \, .
\end{equation}
Here $G^{ab}$ is the inverse of the metric in $\beta$-space $G_{ab}$ defined by $G_{ab} \, \dot\beta^a \, \dot\beta^b \equiv \sum (\dot\beta^a)^2 - (\sum \dot\beta^a)^2$. The metric $G_{ab}$ (which also defines the kinetic term of the $\beta$'s), has signature $-++$ and plays a crucial r\^ole in our problem. 
The fermionic quantization conditions (\ref{eq3}) amount to saying that the $3 \times 4 = 12$ redefined gravitino operators $\hat\Phi_A^a$ constitute a Clifford algebra in a 12-dimensional space [with signature $(+^8 , -^4)$]. The quantization of the gravitino field is then obtained by representing the twelve $\Phi$'s as $64 \times 64$ ``gamma matrices'' which act on a 64-dimensional ``spinorial'' wave function of the universe, say $\Psi (\beta^a , \varphi^a)$. The constraints associated to the Lagrange multipliers in Eq.~(\ref{eq2}) are then imposed \`a la Dirac as conditions on the state $\Psi$:
\begin{equation}
\label{eq4}
\hat{\mathcal S}_A \, \Psi = 0 \, , \quad \hat H \, \Psi = 0 \, , \quad \hat H_a \, \Psi = 0 \, .
\end{equation}

As in the spin-$\frac12$ case~\cite{Damour:2011yk}, we find that the three diffeomorphism constraints are equivalent to requiring $\hat p_{w^a} \Psi (\beta , \varphi) = 0$, i.e. that the wave function $\Psi$ does not depend on the three Euler angles $\varphi^a$. As the $\varphi$'s do not appear in the other constraints, we are left with finding a spinorial wave function $\Psi (\beta^a)$ satisfying the four supersymmetry constraints together with the Hamiltonian one. This raises the usual issue of whether, starting from the classical expressions for ${\mathcal S}_A$ and $H$, one can define an ordering such that their quantum versions $\hat{\mathcal S}_A$, $\hat H$ satisfy an algebra which consistently closes so as to allow for the existence of states satisfying  Eqs.~(\ref{eq4}). One of the crucial results of our work is that we have explicitly verified that this is the case.

Specifically, requiring that the ``real'' (i.e. Majorana) classical ${\mathcal S}_A$ be quantized so as to satisfy the same hermiticity condition, say $\hat{\mathcal S}_A^{\dagger} = \hat{\mathcal S}_A$, than the $\hat\Phi$ operators  they are built from $(\hat\Phi_A^{a\dagger} = \hat\Phi_A^a)$, determines a {\it unique} ordering, of the form
\begin{eqnarray}
\label{eq5}
\hat{\mathcal S}_A &= &-\frac12 \sum_a \hat\pi_a \, \Phi_A^a + \frac12 \sum_a e^{-2\beta^a} (\gamma^5 \, \Phi^a)_A \\
&- &\frac18 \coth \beta_{12} (\hat S_{12} (\gamma^{\hat1 \hat2} \, \hat\Phi^{12})_A + (\gamma^{\hat1 \hat2} \, \hat\Phi^{12})_A \, \hat S_{12}) \nonumber \\
&+ &{\rm cyclic}_{(123)} + \frac12 (\hat{\mathcal S}_{A}^{\rm cubic} + \hat{\mathcal S}_A^{{\rm cubic} \, \dagger}) \, , \nonumber 
\end{eqnarray}
where $\gamma^5 := \gamma^{\hat 0\hat 1\hat 2\hat 3}$, $\beta_{12} := \beta^1-\beta^2$, $\hat\Phi^{12} := \hat\Phi^1 - \hat\Phi^2$,
\begin{eqnarray}
\label{eq6}
\hat S_{12} (\hat\Phi) &= &\frac12 [(\bar{\hat\Phi}^3 \, \gamma^{\hat 0\hat 1\hat 2} (\hat\Phi^1 + \hat\Phi^2)) + (\bar{\hat\Phi}^1 \, \gamma^{\hat 0\hat 1\hat 2} \, \hat\Phi^1) \\
&+ &(\bar{\hat\Phi}^2 \, \gamma^{\hat 0\hat 1\hat 2} \, \hat\Phi^2) - (\bar{\hat\Phi}^1 \, \gamma^{\hat 0\hat 1\hat 2} \, \hat\Phi^2)] \, , \nonumber
\end{eqnarray}

\begin{eqnarray}
\hat{\mathcal S}_A^{\rm cubic} &= &\frac14 \sum_a (\bar{\hat\Psi}_{0'} \, \gamma^{\hat 0} \, \hat\Psi_{\hat a}) \, (\gamma^{\hat 0} \, \hat\Psi_{\hat a})^A - \frac18 \sum_{a,b} (\bar{\hat\Psi}_{\hat a} \, \gamma^{\hat 0} \, \hat\Psi_{\hat b}) \, (\gamma^{\hat a} \, \hat\Psi_{\hat b})^A \nonumber \\
&+ &\frac18 \sum_{a,b} (\bar{\hat\Psi}_{0'} \, \gamma^{\hat a} \, \Psi_{\hat b}) ((\gamma^{\hat a} \, \Psi_{\hat b})^A + (\gamma^{\hat b} \, \Psi_{\hat a})^A) \, , \nonumber
\end{eqnarray}
with $\hat\Psi_{0'} := \gamma_{\hat 0} \, \sum_a \gamma^{\hat a} \, \hat\Psi_{\hat a}$. We then proved that this unique, hermitian ordering of $\hat{\mathcal S}_A$ defines a corresponding unique ordering of the quantum Hamiltonian $\hat H$ such that the four $\hat{\mathcal S}_A$'s satisfy a (super)algebra of the form $\hat{\mathcal S}_A \, \hat{\mathcal S}_B + \hat{\mathcal S}_B \, \hat{\mathcal S}_A = 4 \, i \, {\sum_C} \, \hat L_{AB}^C (\beta) \, \hat{\mathcal S}_C + \frac12 \, \hat H \delta_{AB}$. Such an algebra (with $\hat{\mathcal S}_C$ on the right of $\hat L_{AB}^C$),  further implies that the commutator $[\hat{\mathcal S}_A , \hat H]$ closes on the $\hat{\mathcal S}_A$'s and $\hat H$, and is nicely compatible with the Dirac quantization of the constraints. We found the following explicit form of $\hat H$ (here written after elimination of the angles $\varphi^a$'s)
\begin{equation}
\label{eq7}
2 \, \hat H = G^{ab} (\hat\pi_a + i \, A_a)(\hat\pi_b + i \, A_b) + \hat\mu^2 + \hat W (\beta) \, ,
\end{equation}
where $\hat\pi_a = -i \, \partial_a$ (with $\partial_a := \partial / \partial \beta^a$), and the ``vector potential'' $A_a$ is a pure gradient: $A_a = \partial_a \, \ln \, F$ with $F = e^{\frac34 \, \beta^0} (\sinh \beta_{12} \, \sinh \beta_{23} \, \sinh \beta_{31} )^{-1/8}$. We separated the ``potential term'' in the Wheeler-DeWitt-(WDW)-type equation (\ref{eq7}) into two parts: (i) the $\beta$-independent operator $\hat\mu^2$, which plays the r\^ole of a spin-dependent ``squared-mass'' operator in the Klein-Gordon-like equation (\ref{eq7}), and (ii) the $\beta$-dependent (and spin-dependent) operator $\hat W (\beta)$ whose meaning will be discussed next.  Note that, as the vector potential $A_a$ in equation   (\ref{eq7}) is a pure gradient, $A_a = \partial_a \, \ln \, F$, it can be eliminated, without changing the other terms, by working with the rescaled wave function $\Psi' (\beta) = F(\beta)^{-1} \Psi (\beta)$, i.e. $2 \, F^{-1} \, \hat H (F \Psi') =  (G^{ab} \hat\pi_a  \hat\pi_b  + \hat\mu^2 + \hat W (\beta)) \Psi'$ .

One of the main results of this work concerns the Kac-Moody structures hidden in the (exact) quantum Hamiltonian (\ref{eq7}). First, let us recall that the wave function of the universe $\Psi (\beta)$ is (in view of Eq.~(\ref{eq3})) a 64-component spinor of ${\rm Spin} (8,4)$ which depends on the three logarithmic scale factors $\beta^1 , \beta^2 , \beta^3$. In other words, supergravity describes a Bianchi IX universe as a relativistic {\it spinning particle} moving in $\beta$-space. The spinorial wave function $\Psi (\beta)$ must satisfy four separate Dirac-like equations $\hat{\mathcal S}_A \Psi = \left( + \frac{i}2 \, \Phi_A^a \partial_a + \ldots \right) \Psi = 0$ (where the $\Phi_A^a$'s are four separate triplets of $64 \times 64$ gamma matrices). As shown above, these first-order Dirac-like equations imply that $\Psi$ necessarily satisfy the second-order, Klein-Gordon-like equation $\hat H \Psi = \left(-\frac12 \, G^{ab} \partial_a \partial_b + \ldots \right) \Psi = 0$. The first basic Kac-Moody feature hidden in this dynamics of the universe is the fact that the (Lorentzian-signature) metric $G_{ab}$ defining the kinetic term of the ``$\beta$-particle'' is the metric in the Cartan subalgebra of the hyperbolic Kac-Moody algebra $AE_3$~\cite{Damour:2001sa}. Next, we find that the potential term $\hat W (\beta)$ in Eq.~(\ref{eq7}) is naturally decomposed into three different pieces which all carry a deep Kac-Moody meaning. Namely, we have
\begin{equation}
\label{eq8}
\hat W (\beta) = W_g^{\rm bos} (\beta) + \hat W_g^{\rm spin} (\beta) + \hat W_{\rm sym}^{\rm spin} (\beta) \, .
\end{equation}
Here,
\begin{equation}
\label{eq9}
W_g^{\rm bos} (\beta) = \frac12 \, e^{-4\beta^1} - e^{-2 (\beta^2 + \beta^3)} + {\rm cyclic}_{123}
\end{equation}
is the well-known bosonic potential describing the usual dynamics of  Bianchi IX oscillations~\cite{Belinsky:1970ew,Misner:1969hg}. Its Kac-Moody meaning is that it is constructed from Toda-like exponential potentials $\sim e^{-2\alpha_{ab}(\beta)}$ involving the following six linear forms in the $\beta$'s: $\alpha_{ab}^g (\beta) := \beta^a + \beta^b$, $a,b = 1,2,3$. These six linear forms coincide with the six roots of $AE_3$ located at level $\ell = 1$ (``gravitational walls'', linked to the level-$1$ $AE_3$ ``dual-graviton'' coset field $\phi_{ab} = \phi_{ba}$ of Ref.~\cite{Damour:2002et}). The purely bosonic (spin-independent) potential $W_g^{\rm bos} (\beta)$ is accompanied, in supergravity, by a spin-dependent complementary piece of the form
\begin{eqnarray}
\label{eq10}
\hat W_g^{\rm spin} (\beta , \hat\Phi) &= &e^{-\alpha_{11}^g (\beta)}  \hat J_{11} (\hat\Phi) + e^{-\alpha_{22}^g (\beta)} \hat J_{22} (\hat\Phi) \nonumber \\
&+ &e^{-\alpha_{33}^g (\beta)} \hat J_{33} (\hat\Phi) \, . 
\end{eqnarray}
This involves the three dominant (gravitational) Kac-Moody roots $\alpha_{11}^g (\beta) = 2\beta^1$, etc. each one being coupled to an operator that is {\it quadratic} in the gravitino variables, namely (modulo cyclic permutations)
\begin{equation}
\label{eq11}
\hat J_{11} (\hat\Phi) = \frac12 \, [\bar{\hat{ \ \, \Phi^1}} \gamma^{\hat 1\hat 2\hat 3} (4 \hat\Phi^1 + \hat\Phi^2 + \hat\Phi^3) + \bar{\hat{ \ \, \Phi^2}} \, \gamma^{\hat 1\hat 2\hat 3} \, \hat\Phi^3] \, .
\end{equation}
The third contribution to $\hat W(\beta)$ involves the three level-$0$ Kac-Moody roots $\alpha_{12}^{\rm sym} (\beta) := \beta^1 - \beta^2 \equiv \beta_{12}$, $\alpha_{23}^{\rm sym} (\beta) := \beta^2 - \beta^3$, $\alpha_{31}^{\rm sym} := \beta^3 - \beta^1$ (``symmetry walls''); each one being coupled to an operator that is {\it quartic} in the $\hat\Phi$'s, namely
\begin{equation}
\label{eq12}
\hat W_{\rm sym}^{\rm spin} (\beta) = \frac12 \ \frac{(\hat S_{12} (\hat\Phi))^2 - 1}{\sinh^2 \alpha_{12}^{\rm sym} (\beta)} + {\rm cyclic}_{123} \, ,
\end{equation}
where the {\it spinor operators} $\hat S_{12} (\hat\Phi)$, whose squares enter Eq.~(\ref{eq12}), are exactly those defined in Eq.~(\ref{eq6}) above, which entered the $\hat{\mathcal S}$'s.

A truly remarkable fact, which clearly shows the hidden r\^ole of Kac-Moody structures in supergravity, is that the operators entering $\hat H$ as (spin-dependent) basic blocks, $\hat S_{12}, \hat S_{23} , \hat S_{31}, \hat J_{11} , \hat J_{22} , \hat J_{33}$ generate (via commutators) a Lie-algebra which is a 64-dimensional representation of the (infinite-dimensional) ``maximally compact'' sub-algebra, $K(AE_3)$, of $AE_3$. Indeed, the $\hat S$'s generate the $(\ell = 0)$ sub-algebra $SO(3)$ of $K(AE_3)$ ($[\hat S_{12} , \hat S_{23}] = + i \, \hat S_{31}$, etc.), while we have checked that the gravitational generator $\hat J_{11}$ can be identified with the crucial level-1 Lie-algebra element denoted $J_{\alpha_*} = E_{\alpha_*} - E_{-\alpha_*}$ in Ref.~\cite{Damour:2009zc}. More precisely, we found that the generators $\hat S_{ab}$ ($a < b$), and $\hat J_{ab}$ (with, e.g., $\hat J_{12} := -\frac{i}2 \, [\hat S_{12} , \hat J_{11}]$, etc.) are {\it second-quantized} versions of the (first-quantized) level-$0$ and level-$1$ $K(AE_3)$ generators defining the 12 dimensional vector-spinor representation of $K(AE_3)$~\cite{Damour:2005zs,de Buyl:2005mt,Damour:2009zc}. [This means that their quantum commutators with the gravitino operators $\hat\Psi_{\hat a}^A$ reproduce the Lie-algebra-bracket actions of $J_{[ab]}^{(\ell = 0)}$ and $J_{(ab)}^{(\ell = 1)}$ on a ``classical'' vector-spinor gravitino $\Psi_{\hat a}^A$.]

Finally, let us consider the $\beta$-independent, operator-valued squared-mass contribution $\hat\mu^2$ to the Hamiltonian (\ref{eq7}). This term gathers many complicated, quartic-in-fermions contributions (including the infamous $\psi^4$ terms in the original, second-order supergravity action). However, at the end of the day two remarkable (Kac-Moody-related) facts emerge: (i) $\hat\mu^2$ belongs to the {\it center} of the algebra generated by the $K(AE_3)$ generators $\hat S_{ab} , \hat J_{ab}$ (i.e. it commutes with all of them), and (ii) the quartic operator $\hat\mu^2$ can be expressed in terms of the square of a very simple operator (which also commutes with $\hat S_{ab} , \hat J_{ab}$), namely, we find
\begin{equation}
\label{eq13}
\hat\mu^2 = \frac12 - \frac78 \, \hat C_F^2
\end{equation}
where $\hat C_F := \frac12 \, G_{ab} \bar{\hat{ \ \, \Phi^a}} \, \gamma^{\hat 1\hat 2\hat 3} \, \hat\Phi^b$. As we shall discuss next, $\hat C_F$ is related to the fermion number operator $\hat N_F$ by $\hat C_F \equiv \hat N_F - 3$.

So far we have presented some of the main formal results about our new way of quantizing the supersymmetric squashed three-sphere, and their relation to Kac-Moody structures. In addition, we succeeded in controlling in detail the space of solutions of this model. Let us briefly sketch our results. To do so (and to connect our results to previous, partial results on the same model), it is useful to combine the (hermitian) operators $\hat\Phi_A^a$ into fermionic annihilation and creation operators; $b_+^a = \hat\Phi_1^a + i \, \hat\Phi_2^a$, $b_-^a= \hat\Phi_3^a - i \, \hat\Phi_4^a$, $b_+^{a\dagger} = \hat\Phi_1^a - i \, \hat\Phi_2^a$, $b_-^{a\dagger} = \hat\Phi_3^a + i \, \hat\Phi_4^a$, which satisfy $\{ b_{\sigma}^a , b_{\sigma'}^{b\dagger} \} = 2 \, G^{ab} \delta_{\sigma\sigma'}$. The 64 states of ${\rm Spin} (8,4)$ can then be constructed from the empty state $\vert 0 \rangle_-$ (annihilated by the six $b_{\sigma}^a$'s) by acting with a certain number of $b_{\sigma}^{a\dagger}$ operators. Actually, $\hat N_F = \hat C_F + 3$ counts this number of $b^{\dagger}$ operators. $\hat N_F$ commutes with $\hat H$ (though not with $\hat{\mathcal S}$) and solutions can be searched for at each fermionic level. We found the following results for the complete space of solutions, say ${\mathcal V}^{(N_F)}$, of $\hat{\mathcal S}_A \Psi = 0$ at level $N_F$ (i.e. $\hat N_F \Psi = N_F \Psi$): ${\mathcal V}^{(0)} = V_1^{(0)}$ is one-dimensional; ${\mathcal V}^{(1)} = V_2^{(1)}$ is two-dimensional; ${\mathcal V}^{(2)} = V_3^{(2)} \oplus V_{1,\infty^2}^{(2)}$ is the direct sum of a three-dimensional space $V_3^{(2)}$ and of an infinite-dimensional space $V_{1,\infty^2}^{(2)}$ parametrized by one constant and two (complex) functions of two (real) variables; ${\mathcal V}^{(3)} = V_{2,\infty^2}^{(3)} \oplus V_{2,\infty^2}^{(3)}$ is the direct sum of two infinite-dimensional spaces, each one of which involves as free data two parameters and two functions of two variables. Moreover, when $4 \leq N_F \leq 6$, there is a duality under which ${\mathcal V}^{(N_F)}$ is one-to-one mapped to ${\mathcal V}^{(6-N_F)}$. Our results significantly differ from the conclusions of previous works. [One should, however, keep in mind that our quantization scheme is somewhat different from the ones used before.] The most striking disagreement is that all previous authors~\cite{D'Eath:1993ki,D'Eath:1993up,Csordas:1995kd,Csordas:1995qy,Obregon:1998hb} agreed on the inexistence of solutions when $N_F$ is odd, while we proved the existence of solutions for $N_F = 1,3$ and $5$. For instance at $N_F = 1$ we found a two-dimensional space of solutions of the form $\sum_{\sigma , a} \, C_{\sigma} f_a (\beta) \, b_{\sigma}^{a\dagger} \, \vert 0_- \rangle$, ($\sigma = \pm$) where we could compute the explicit form of the three functions $f_a (\beta)$. Even at $N_F = 2$ and $4$, where we partially confirm the claim of~\cite{Csordas:1995kd,Csordas:1995qy} about the existence of solutions parametrized by the same amount of initial data as a Klein-Gordon (or WDW) equation, we found extra, discrete solutions. Moreover, at $N_F = 0$ and $6$, where our results qualitatively agree with previous ones, we find some significant differences coming from our treatment of the diffeomorphism constraint. E.g. the unique ``ground state'' at $N_F = 0$ reads
\begin{eqnarray}
\textstyle \Psi^{(0)} &= &\textstyle \exp \left( - \frac74 \, \beta^0 \right) (\sinh \beta_{12} \sinh \beta_{23} \sinh \beta_{31})^{3/8} \nonumber \\
&&\textstyle\exp \left( -\frac12 \sum_a \exp (-2 \, \beta^a) \right) \vert 0 \rangle_- \, , \nonumber
\end{eqnarray}
which differs from previous results, notably by the effect of $\sinh \beta_{ab}$ factors vanishing on the three symmetry walls.

Finally, our results allow us to qualitatively describe the structure of the general solution (belonging to the infinite-dimensional pieces of ${\mathcal V}^{(2)} , {\mathcal V}^{(3)}$ and ${\mathcal V}^{(4)}$) near a cosmological singularity. First, in the intermediate asymptotics where $\beta^0 = \beta^1 + \beta^2 + \beta^3$ (which measures the {\it cologarithm} of the volume of the universe) is large but not too large, we can qualitatively describe the evolution of the state $\Psi (\beta)$ as a {\it quantum fermionic billiard}. The spinning $\beta$-particle undergoes a sequence of quantum reflections on the gravitational and/or symmetry potential walls that appear both in the $\hat{\mathcal S}$'s, Eq.~(\ref{eq5}), and in $\hat H$, Eq.~(\ref{eq7}). As in the Grassmannian case~\cite{Damour:2009zc}, and in the spin-$\frac12$ toy problem studied in~\cite{Damour:2011yk}, we were able to show that the reflections on the various walls are given, in the small wavelength limit, by operators of the form $\exp \left( -i \, \frac{\pi}2 \, \hat\varepsilon_{\alpha_{ab}} \hat S_{\alpha_{ab}}\right)$ for symmetry walls (with $\hat\varepsilon_{\alpha_{ab}}^2 = 1$), and $\exp\left( -i \, \frac{\pi}2 \, \hat J_{aa} \right)$ for the dominant gravitational walls. This exhibits again a Kac-Moody structure: the (small-wavelength) quantum reflections generate a {\it spinorial extension of the Weyl group} of $AE_3$. On the other hand, in the asymptotic regime where $\beta^0 \to +\infty$ (i.e. formally, for infinitely small volumes) the qualitative dynamics might become essentially monitored by the {\it sign} of the eigenvalues of the squared-mass operator $\hat\mu^2$. Indeed, in this limit the billiard walls become more and more separated, so that the $\beta$-particle spends more and more ``$\beta^0$-time'' far from the walls, i.e. in a domain where $\hat W (\beta) \ll \hat\mu^2$ in Eq.~(\ref{eq7}). The simple result (\ref{eq13}) then suggests that the three generic-data components (at levels $N_F = 2,3,4$, i.e. $C_F = -1,0,+1$) of the wave function of the universe might have very different asymptotic behaviors near the singularity. Indeed, when $N_F = 3$, $C_F = 0$, $\mu^2 = \frac12$ is strictly positive so that the corresponding piece of $\Psi (\beta)$ might behave like an ordinary massive particle (with an ultimate behavior which oscillates in $\beta^0$, or, better in $\rho = \sqrt{-G_{ab} \, \beta^a \beta^b}$~\cite{Damour:2002et,Kleinschmidt:2009cv}, with some power-law decay). By contrast, when $N_F = 2$ ($C_F = -1$) or $N_F = 4$ ($C_F = +1$), $\mu^2 = -\frac38$ is strictly negative so that the corresponding piece of $\Psi (\beta)$ might behave like a tachyon. We leave to future work a discussion of the possible physical implications of these behaviors. 
 Let us only note here that, contrary to the spin-$\frac12$ case (or to pure gravity), where quantization generically allows for arbitrary ordering constants in the WDW equation, supergravity (together with a natural hermiticity requirement) has uniquely fixed all ordering constants in $\hat{\mathcal S}_A$, and thereby in $\hat H$. This suggests that one should seriously consider the implication (never suggested before) of having a tachyonic $(\mu^2 < 0)$ behavior of part of the wave function of the universe near the singularity (located at $\beta^0=+\infty$).  Classically, $\mu^2 < 0$ would ensure an ultra-chaotic behavior; quantum mechanically, it allows one to impose the boundary condition that $\Psi (\beta)$ vanishes exponentially at the singularity.


\smallskip\noindent\emph{Acknowledgments.~} We thank  G. Bossard,  P. Deligne, V. Kac, A. Kleinschmidt, V. Moncrief, H. Nicolai, and K. Stelle  for informative discussions. Ph. S. thanks IHES for its kind hospitality; his work has been partially supported by ``Communaut\'e fran\c caise de Belgique -- Actions de Recherche concert\'ees'' and by IISN-Belgium (convention 4.4511.06).

\end{document}